%% file: paper.tex
\documentclass[sigconf]{acmart}
\usepackage[ruled, noline, linesnumbered, noend]{algorithm2e}

\copyrightyear{2026}
\acmYear{2026}
\setcopyright{cc}
\setcctype[4.0]{by}
\acmConference[QuNet '26]{3rd ACM SIGCOMM Workshop on Quantum Networks and Distributed Quantum Computing}{August 17--21, 2026}{Denver, CO, USA}
\acmBooktitle{3rd ACM SIGCOMM Workshop on Quantum Networks and Distributed Quantum Computing (QuNet '26), August 17--21, 2026, Denver, CO, USA}
\acmDOI{10.1145/3833409.3833449}
\acmISBN{979-8-4007-2892-1/2026/08}

\begin{document}

\title[DPRQ: A Dynamic Programming-based Qubit Routing Algorithm]{DPRQ: A Dynamic Programming-based Qubit Routing Algorithm for Collective Communication in Distributed Quantum Computing}

\author{Dhaval Vaidya}
\email{dkvaidya@ncsu.edu}
\orcid{0009-0003-5372-7246}
\affiliation{%
  \institution{North Carolina State University}
  \city{Raleigh} 
  \state{North Carolina}
  \country{USA}
}

\author{Ruozhou Yu}
\email{ryu5@ncsu.edu}
\orcid{0000-0003-0905-5158}
\affiliation{%
  \institution{North Carolina State University}
  \city{Raleigh} 
  \state{North Carolina} 
  \country{USA}
}

\begin{abstract}
    Distributed quantum computing (DQC) offers a promising approach to scale quantum computing by overcoming the resource limitations of a single quantum processor. However, inter-node communication remains a major bottleneck of DQC due to inefficient and error-prone entanglement distribution. Optimizing inter-node communication can not only reduce the amount of entanglement resource needed to execute a quantum circuit but also improve execution speed and accuracy of the results. This paper proposes DPRQ, a qubit routing algorithm for minimizing inter-node communication in distributed quantum circuits divided into collective communication blocks. Unlike current approaches that utilize greedy block-level qubit routing strategies, DPRQ employs a dynamic programming-based technique focused on global circuit-level optimization, while capturing inter-block dependencies. We evaluated DPRQ on four sets of quantum circuits and a variety of DQC configurations. The results demonstrate that DPRQ's innovative routing strategy achieves an average of 24.40\% reduction with a maximum of 85.06\% reduction in inter-node communication, when compared to the state-of-the-art collective communication-based DQC compiler QuComm.
\end{abstract}



\maketitle

\input{introduction}
\input{background}
\input{techniques}
\input{results}

\input{conclusion}

\bibliographystyle{ACM-Reference-Format}
\bibliography{reference}

\end{document}

%% file: introduction.tex
\section{Introduction}
\label{sec:introduction}
In the NISQ era, quantum processors are erroneous and cannot accommodate enough qubits to be fully fault-tolerant. \textbf{Distributed Quantum Computing (DQC) is the technique of using a network of multiple quantum processors to perform a task.} It enhances the scalability of quantum computing by helping surpass the qubit capacity constraints of a single quantum processor.

Different techniques of DQC are being investigated around the world, including those in hardware design \cite{Daiss2021, Hermans2022, Pompili2021, Ruf2021} as well as program-level compilation \cite{Andrs-Martnez2019, 10.1098/rspa.2012.0686, Dadkhah2022, Daei2020, Davarzani2020, DiAdamo2021, Ferrari2021, H_ner_2021, Zomorodi-Moghadam2018, wu2022autocommframeworkenablingefficient, 10.1145/3613424.3614253, Chatterjee2022}. In the general model of DQC \cite{Andrs-Martnez2019, 10.1098/rspa.2012.0686, Dadkhah2022, Daei2020, Davarzani2020, DiAdamo2021, Ferrari2021, H_ner_2021, Zomorodi-Moghadam2018, wu2022autocommframeworkenablingefficient, 10.1145/3613424.3614253}, one Einstein-Podolsky-Rosen (EPR) pair is consumed to perform one inter-node communication. However, the error-prone nature of the generation and maintenance of EPR pairs makes it a scarce quantum resource and inter-node communication; an expensive process \cite{Pompili2021}. To increase the efficiency of inter-node communication, some approaches focus on reducing inter-node communication using innovative compiling techniques such as \emph{burst communication} and \emph{collective communication} \cite{10.1145/3613424.3614253, wu2022autocommframeworkenablingefficient}.

Among various approaches, \emph{collective communication} is a novel approach to optimize inter-node communication, pioneered in the seminal work QuComm \cite{10.1145/3613424.3614253}. QuComm devises a strategy to optimize node-to-node communication by dividing a quantum circuit into collective communication blocks and optimizing the routing within each block to reduce EPR consumption.

A major limitation of QuComm is the greedy routing mechanism of inter-node gates that solely considers routing optimization within each collective communication block, significantly limiting the reduction in EPR consumption across collective communication blocks. We present DPRQ, a dynamic programming-based qubit routing algorithm aimed at minimizing the number of EPR pairs required for inter-node communication in a collective communication setting. Retaining the logic of forming collective communication blocks as \cite{10.1145/3613424.3614253}, DPRQ uses an innovative dynamic programming-based approach to create a routing strategy that intends to minimize inter-node communication across an entire DQC circuit. The results demonstrate that DPRQ offers an average reduction of 24.40\% in inter-node communication with a maximum reduction of 85.06\% when compared to the baseline. The primary contributions of our paper are summarized as follows:
\begin{itemize}
    \item We create an intelligent qubit routing framework specifically aimed at reducing the EPR costs in a DQC circuit divided into collective communication blocks.
    \item We evaluate the algorithm against the state-of-the art baseline and prove its effectiveness in reducing inter-node communication.
\end{itemize}

The remainder of this paper is structured as follows. In Section \ref{sec:background}, we provide a comprehensive background of the types of qubits in a quantum processor, the TP-Comm quantum communication protocol, collective quantum communication, and the \emph{communication fusion} stage of \emph{QuComm}~\cite{10.1145/3613424.3614253} where the circuit is divided into collective communication blocks. In Section \ref{sec:techniques}, we introduce the technical innovation of DPRQ for efficient routing in a collective quantum communication setup. In Section \ref{sec:results}, we evaluate DPRQ on four different quantum circuits and demonstrate the performance of DPRQ for a variety of different circuit configurations. In Section \ref{sec:conclusion}, we conclude the paper.

%% file: background.tex
\section{Background}
\label{sec:background}

\subsection{Qubits, Qubit Layout and TP-Comm} In a quantum processor, the physical qubits that can establish a remote entanglement are called \emph{communication qubits}, whereas the physical qubits that store program information are known as \emph{data qubits}. A communication qubit can be used to store program information; however, a data qubit cannot be used to establish a remote entanglement. Each node has a fixed number of communication qubits, which is known as the \emph{EPR capacity} of the node \cite{10.1145/3613424.3614253}. \emph{Logical qubits} are computational abstractions that represent qubits in a quantum circuit. 

At any given time $t$, the \emph{qubit layout} describes the mapping of each logical qubit in the circuit to the corresponding node where that qubit currently occupies a physical qubit.

The TP-Comm protocol uses quantum teleportation to transfer a qubit from one node to another using a remote EPR pair. One invocation of TP-Comm consumes one EPR pair and transfers one qubit to another node. As a qubit is relocated to another node, the qubit layout of the DQC network changes after an invocation of TP-Comm. 

\subsection{Collective Communication for DQC}
Consider a qubit interaction graph $G_q = (V_q, E_q)$ where vertices represent qubits and edges represent gates. A connected $G_q$ having more than two vertices implies that certain qubits (distributed over various compute nodes) are involved in the execution of more than one gate. If the gates involved in $G_q$ are executed collectively at a common node, it can lead to a reduction in inter-node communication. This is known as collective communication. Formally, a \emph{collective communication block} is a group of inter-node gates having a connected inter-node interaction graph on qubits over multiple nodes \cite{10.1145/3613424.3614253}. 

\subsection{Communication Fusion}
\emph{Communication fusion} is the preprocessing stage in which the circuit is inspected for opportunities for collective communication and systematically divided into various collective communication blocks. This stage is reproduced from the description given by \cite{10.1145/3613424.3614253}.

We assume that the circuit consists of one-qubit and two-qubit gates. Initially, the first non-local gate creates a new block, and each gate in the circuit having overlapping qubits with the block is subsequently iterated. A non-local gate is fused into the block if the estimated routing cost denoted by \eqref{eq:cost-fusion} in the block does not increase after fusion, while a local gate is fused if the implementation cost of the block after fusion remains unaffected. When no gate could be merged in the block without increasing post-fusion cost estimate, a new block is formed with the next non-local gate, and the entire process is repeated until no non-local gates in the circuit remain unassigned. The remaining unassigned local gates form individual blocks, which is a design choice made by us to ensure the assignment of all gates.

Assume that $E(n_a)$ is the EPR capacity of node $n_a$, the number of qubits involved in block $b_0$ is $H(b_0)$, and $H(b_0 + b_1 - n_a)$ is the total number of qubits involved in blocks $b_0$ and $b_1$ but not present in node $n_a$. If a block $b_1$ is to be fused into block $b_0$, then the estimated inter-node cost of executing blocks $b_0$ and $b_1$ is given as $F(b_0, b_1)$ in Equation \eqref{eq:cost-fusion}.

{
\begin{align}
F(b_0, b_1) &= \min_{n_a \in \text{nodes}}
\bigg\{
    \max\!\left\{ 2 \cdot \big(H(b_0 + b_1 \text{---} n_a) \text{---} E(n_a)\big),\, 0 \right\} \nonumber \\
    &\quad +
    \min\!\left\{ E(n_a),\, H(b_0 + b_1 \text{---} n_a) \right\}
\bigg\}.
\label{eq:cost-fusion}
\end{align}
}

The reasoning is that if $H(b_0 + b_1 \text{---} n_a) \leq E(n_a)$, then all the qubits in block $b_0$ and $b_1$ can be sent to $n_a$ using one EPR pair for each qubit. However, if $H(b_0 + b_1 \text{---} n_a) > E(n_a)$, this means that the number of qubits to be transferred to $n_a$ is more than the availability of communication qubits at $n_a$. Each of the additional qubits has to be swapped into node $n_a$ (with another qubit already present at $n_a$) using an inter-node SWAP gate that uses 2 invocations of TP-Comm. This is the reason why there is a multiplying factor of 2 in the first term of \eqref{eq:cost-fusion}.

%% file: techniques.tex
\section{Techniques}
\label{sec:techniques}
We present DPRQ, a dynamic programming-based algorithm for efficient routing of qubits in a collective communication setup. DPRQ relies on two key processes for optimized qubit routing: (1) Intra-block communication cost, and (2) DP-based inter-block qubit routing.

\begin{algorithm}[!t]
\caption{Procedure to calculate intra-block routing cost.}
\label{alg:routing-cost}
    \SetKwInOut{Input}{Input}
    \SetKwInOut{Output}{Output}
    \SetKwProg{Fn}{Function}{ :}{end}
    \Input{Current block $b$, aggregator node $n_a$,\\ current qubit layout $l$, \\ EPR capacity $e$\;}
    \Output{Final qubit layout $l$,\\ final cost $T$\;}
    \Fn{\textsc{IntraBlockRouting}($b$, $n_a$, $l$, $e$)}{
    $T \gets 0$\;
    \For{gate $g$ in $b$}{
        \If{$g$ is a local gate}{
            continue\;
        }
        $n \gets $ Current node where the first qubit $q$ of $g$ is present\;
        $n' \gets $ Current node where the second qubit $q'$ of $g$ is present\;
        $P_a \gets $ Set of all shortest paths from $n$ to $n_a$\;
        $P_a' \gets $ Set of all shortest paths from $n'$ to $n_a$\;
        
        \For{$p_a$ in $P_a$}{
            \For{$p_a'$ in $P_a'$}{
                $N'' \gets $ Set of all common nodes in $p_a$ and $p_a'$\;
                \For{$n''$ in $N''$}{
                    $c_{n''} \gets 0$\;
                    $c_t \gets $ Cost of sending $q$ to $n''$ via the shortest path\; 
                    $c_{n''} \gets c_{n''} + c_t$\; 
                    $c_t' \gets$ Cost of sending $q'$ to $n''$ via the shortest path\; 
                    $c_{n''} \gets c_{n''} + c_t'$\;
                }
            }
        }
        $n_f \gets $ Node $n''$ for which $c_{n''}$ is minimum\;
        $c_{n_f} \gets$ Cost of sending both qubits $q$ and $q'$ to node $n_f$\;
        $l \gets $ Qubit layout after sending both qubits $q$ and $q'$ to $n_f$\;
        $T \gets T + c_{n_f}$\;
    }
    \textbf{Return} $l, T$\;
    }
\end{algorithm}

\subsection{Intra-block Communication Cost} 

The goal for routing within each collective communication block is to (i) select an aggregator node, and (ii) teleport all the qubits involved in the block to the aggregator node for local execution of inter-node gates. The intra-block communication cost jointly depends on the aggregator node and the teleportation paths of the qubits.

In a uniform DQC network, the EPR cost to transfer a qubit to an aggregator node is the shortest path between the node on which the qubit is present and the aggregator node. However, in a quantum network topology, multiple shortest paths may exist between every pair of nodes. We attempt to pick the shortest path that could lead to the maximum reduction in qubit transfer cost, by leveraging the capability to early-execute some inter-node gates at intermediate nodes. \textbf{This technique of executing gates early on intermediate nodes lying on the shortest qubit transfer paths is called early execution \cite{10.1145/3613424.3614253}}.

We define $T(b_k, n_{k\text{---}1}, n_{k})$ as the inter-node communication cost of routing qubits in a block $b_k$ in \eqref{eq:cost-bk}, where $n_k$ is the aggregator node of the current block $b_k$ and $n_{k\text{---}1}$ is the aggregator node of the previous block $b_{k\text{---}1}$. Additionally, a matrix $L$ is maintained, where $L(b_k, n_k)$ is the final qubit layout after the execution of block $b_k$ with node $n_k$ as the aggregator node. Consider the initial qubit layout before the circuit execution begins as $l''$ and the ordered list of all collective communication blocks as $B$.

Given an aggregator node $n_a$, the current block $b$, the EPR capacity of the nodes $e$, and the initial qubit layout $l$, the \textit{\textsc{IntraBlockRouting}} procedure in Algorithm~\ref{alg:routing-cost} iterates through each two-qubit gate of the block sequentially and finds all potential nodes (including $n_a$) that offer an early execution opportunity for that gate. For each potential node, the EPR cost is calculated to transfer both qubits to these nodes. To calculate the precise routing cost of teleporting a qubit to any specific node using a path, DPRQ sequentially transfers the qubit to each intermediate node and calculates the transfer cost which is one if communication qubits are available (confirmed using the parameter $e$) and two otherwise (using inter-node SWAP gates). Finally, the node offering the least EPR cost is selected, and both qubits are teleported to that node, updating the corresponding qubit layout.

{
\begin{align}
    T(b_k, n_{k\text{---}1}, n_k) =
    \begin{cases}
        \emph{\textsc{IntraBlockRouting}($b_k, n_k, L(b_{k-1}, n_{k-1}), e$)} & \\ \textbf{if } k \in \{1, |B|\text{---}1\}, \\
        \\
        \emph{\textsc{IntraBlockRouting}($b_0, n_0, l'', e$)} & \\ \textbf{if } k = 0.
    \end{cases}
    \label{eq:cost-bk}
\end{align}
}

\subsection{DP-based Inter-block Qubit Routing}

Based on the cost of qubit routing within each block, DPRQ further employs a dynamic programming-based approach to reduce the end-to-end routing cost for the entire circuit, taking into account the resultant qubit layouts after executing each block. For a block $b_k$, DPRQ calculates the cost to select node $n_k$ as the aggregator node for all $n_k\in N$, where $N$ is the set of all nodes in the DQC network. Specifically, DPRQ maintains a matrix of cost $C$, where $C(b_k, n_k)$ denotes the total communication cost to execute all blocks $\{b_0, \dots, b_k\}$ while choosing $n_k$ as the aggregator node for block $b_k$. The cost $C(b_k, n_k)$ is given in \eqref{eq:dp-cost}. 

{
\begin{align}
    C(b_k, n_k) = 
    \begin{cases}
        \min_{\forall n_{k\text{---}1} \in N}\{C(b_{k\text{---}1}, n_{k\text{---}1}) + T(b_k, n_{k\text{---}1}, n_k)\} & \\ \textbf{if } k \in \{1, |B|\text{---}1\},  \\
        \\
    T(b_k, n_{k\text{---}1}, n_k) & \\ \textbf{if } k=0.
    \end{cases}
\label{eq:dp-cost}
\end{align}
}

DPRQ considers all previous aggregator nodes $n_{k\text{---}1} \in N$ for every $n_k \in N$. This ``looking-back'' behavior enables DPRQ to explore a large space of initial qubit layouts for a block $b_k$, since the initial qubit layouts for $b_k$ directly depend on the final qubit layouts after executing $b_{k-1}$. Then it is intuitive that the layout $L(b_{k-1}, n_{k-1})$ corresponds to the minimum cost $C(b_{k-1}, n_{k-1})$. It can be argued that including multiple final qubit layouts for each tuple ($b_{k-1}, n_{k-1}$) might lead to a better optimization, however, it comes at the cost of significant space and time complexity, due to which we include a single best final layout and leave the exploration of storing multiple layouts as future work.

An accurate estimate of cost $T$ denoted by \eqref{eq:cost-bk} gives DPRQ the ability to select a better aggregator node and decrease the number of EPR pairs required for intra-block routing. Additionally, the dynamic programming approach enables DPRQ to optimize routing among consecutive collective communication blocks by considering the best of multiple initial qubit layouts. If $b_x$ is the last block in the circuit, the final number of EPR pairs required to execute the circuit is given as $S(b_x)$ in \eqref{eq:dprq-final-cost}.
\begin{equation}
    S(b_x) = \min_{\forall n_y\in N}C(b_x, n_y).
    \label{eq:dprq-final-cost}
\end{equation}

\subsection{Scalability Analysis}
We analyze the time complexity by referring to Algorithm~\ref{alg:routing-cost} for better understanding. Consider that $N$ is the set of all nodes in a DQC network and $Z$ is the list of all two-qubit gates ordered according to the corresponding collective communication blocks into which the circuit is divided. The shortest paths are calculated in $O(|N| + Y + P \cdot |N|)$ (Lines 8-9), where $Y$ is the number of links in the network and $P$ is the maximum number of shortest paths between two nodes. Common nodes are found in $O(P^2|N|)$ (Line 12). Now, To send a qubit to each unique common node, all shortest paths are calculated again (from qubit to the common node), and the best is chosen. This ensures that when the node is a common node in another set of shortest paths, we can simply ignore it, as we have already calculated the best path. Now, to send qubit to the node, all intermediate nodes are traversed (this is necessary to account for intermediate nodes without available communication qubits)(Lines 13-18). Hence, the complexity for this process is $O(|N|\cdot (|N| + Y + P \cdot |N|))$. The worst-case time complexity of Algorithm~\ref{alg:routing-cost} is $O(|Z| \cdot ((|N| + Y + P \cdot |N|) + P^2|N| + |N|\cdot (|N| + Y + P \cdot |N|))) = O(|Z| \cdot (P^2|N| + P|N|^2 + |N|Y))$. To calculate the routing cost for each node as the aggregator node for a single block, all aggregator nodes of the previous block are considered. Hence, the worst-case time complexity of DPRQ is $O(|Z| \cdot (P^2|N|^3 + P|N|^4 + |N|^3Y))$. Please note that in a large network with rich connectivity, P may be exponential in the size of the network, but for near-term quantum networks, the time complexity is manageable, and for future large-scale networks, heuristic rules can be defined for selection of a polynomial subset of paths, which is a standard practice in classical networks \cite{10.1109/INFOCOM.2018.8486419}.

%% file: results.tex
\section{Results}
\label{sec:results}

\subsection{Experimental Setup}
We have used Python 3.12.8 on a Windows 11 System with 8 CPU cores and 16 GB RAM for simulations. A mesh-grid DQC network topology  \cite{10.1145/3613424.3614253} is employed to evaluate DPRQ on a variety of different DQC circuit configurations. 
Since the qubit topology inside each node is orthogonal to our research direction, no assumptions are made about the same. Furthermore, a nearest-neighbor architecture is adopted, where inter-node communication is restricted to neighboring nodes. It is also presumed that a constant communication channel is established between neighboring nodes.

We evaluated DPRQ on four fault-tolerant quantum circuits that have major real-world applications. They are the Bernstein-Vazirani (BV) circuit (with secret string of all one) \cite{Bernstein1997}, Ripple-Carry Adder (RCA) circuit decomposed into Clifford+T basis \cite{Vedral1996}, fixed EfficientSU2 ansatz for Variational Quantum Eigensolver (VQE) having linear entanglement \cite{Peruzzo2014}, and QAOA (Maxcut Ansatz on a random graph with 10 edges) \cite{farhi2014quantumapproximateoptimizationalgorithm}.

A 95\% confidence interval is also calculated by running 10 experiments for each configuration of QAOA to account for randomness. The BV, RCA, and QAOA circuits are obtained from the open-source implementation of AutoComm \cite{wu2022autocommframeworkenablingefficient}, while the VQE circuit is taken from IBM Qiskit \cite{javadiabhari2024quantumcomputingqiskit}.

\subsection{Baseline, Evaluation Parameters and Evaluation Metric}
As DPRQ uses the notion of collective communication similar to that used by the state-of-the-art QuComm, we used QuComm as the baseline DQC compiler to evaluate the performance of DPRQ. We reproduced the \emph{communication fusion} and use it as the preprocessing stage in DPRQ as well as QuComm. We also reproduced the \emph{communication routing} stage (the qubit routing stage) of QuComm and use it for comparative evaluation with DPRQ's routing algorithm. \footnote{We omit the \emph{communication buffer design} stage in QuComm's reproduction as it is orthogonal to our research direction.} TP-Comm is used as the quantum communication protocol to transfer qubits among nodes. \footnote{Please note that the original QuComm paper uses Cat-Comm as well as TP-Comm; however, we use solely TP-Comm in DPRQ and exclude Cat-Comm from QuComm's reproduction to maintain consistency in analysis. DPRQ can be extended to include Cat-Comm and we leave it for future work.}

In an experimental DQC setup, each node has a limited number of communication qubits. Moreover, a DQC circuit consisting of 150 qubits distributed over 8 compute nodes is physically realizable on certain quantum computing hardware. Hence, we select the following default values of the parameters owing to their practical relevance: EPR capacity = 3, number of data qubits = 150, and number of data qubits per node = 20. Each experiment is performed by keeping the default values for all parameters except one.
The initial logical qubits of the circuit are assigned to the data qubits of node A. Once all the data qubits in node A are occupied, the remaining qubits are assigned to the data qubits of node B, C, etc. following the same logic. The same assignment method is applied to the baseline to ensure fairness in the evaluation. Finally, the number of nodes in the DQC network will be the total number of data qubits divided by the number of data qubits per node.

Inter-node communication relies on remote EPR pairs, and the primary objective of DPRQ is to reduce inter-node communication in a DQC circuit. Hence, the \textbf{number of EPR pairs} (or the number of invocations to TP-Comm) is used as the primary metric to assess the performance of DPRQ.

\begin{figure}[!t]
    \centering
    \includegraphics[width=\columnwidth]{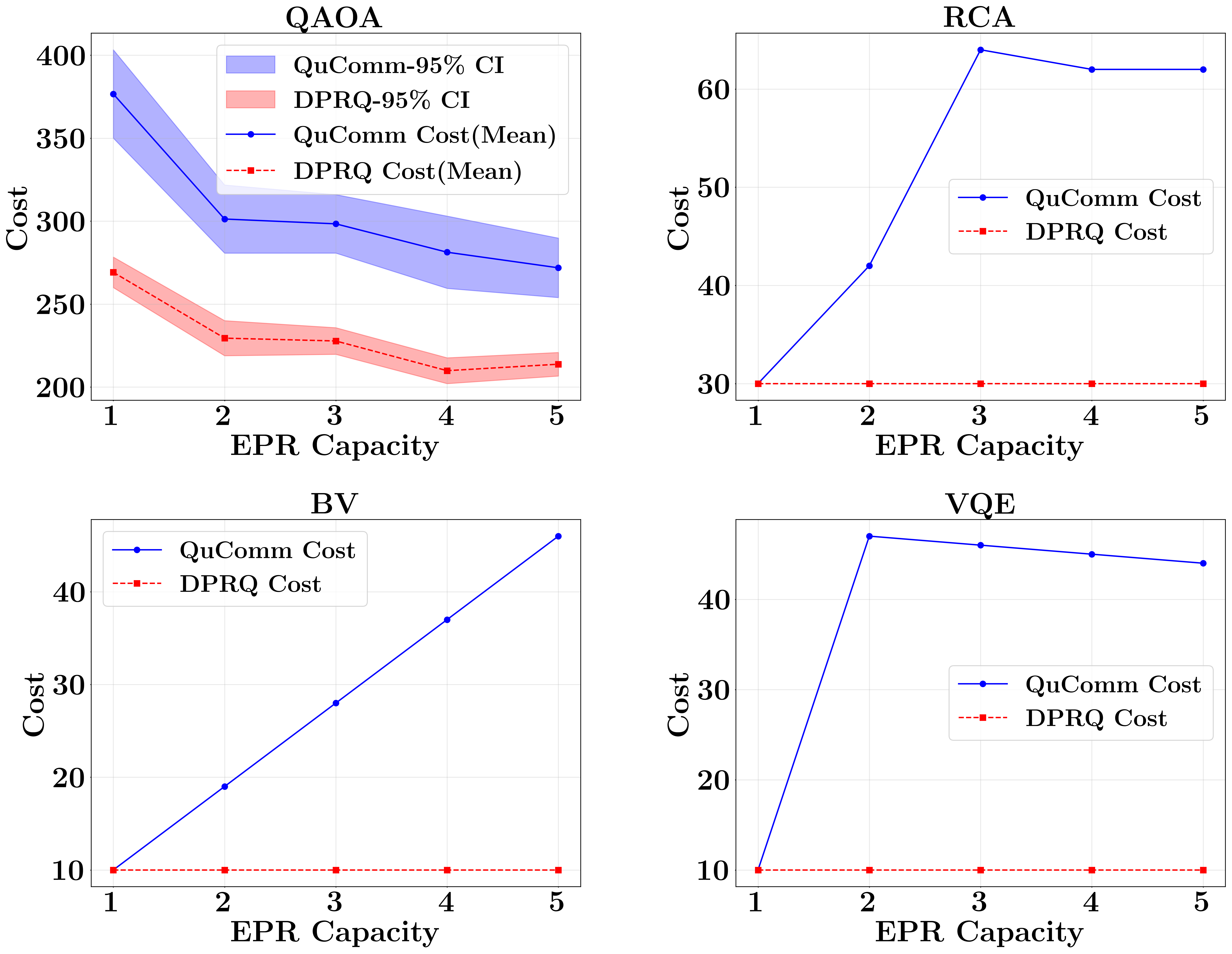}
    \caption{The cost comparison of DPRQ and QuComm on the four benchmark circuits as the EPR capacity increases from 1 to 5. The total number of qubits is 150 and the qubits per node is 20.}
    \label{fig:results-epr}
    \Description{This figure contains four graphs, each containing two line plots. The red line plot shows the EPR cost for DPRQ and blue line plot shows the EPR cost of routing using the baseline as the EPR capacity increases from 1 to 5. The top-left graph displays data for QAOA, the top-right graph displays data for RCA, the bottom-left graph displays data for BV, and the bottom-right graph displays data for VQE. Additionally, in the QAOA graph, there is a 95\% confidence interval shown for both DPRQ and the baseline.}
\end{figure}

\subsection{Effect of Circuit Parameters}
As shown in Figure~\ref{fig:results-epr}, DPRQ consistently achieves lower routing costs for varying EPR capacity values across all four benchmark circuits. DPRQ executes the BV circuit in constant cost owing to its dynamic programming-based approach that enables sequential transfer of a single qubit to each node, avoiding redundant inter-node communication. Similarly, DPRQ executes the VQE and RCA circuits with a near-constant number of EPR pairs using an adaptive routing strategy that remains resilient to change in block patterns. 
The results indicate that DPRQ is advantageous for a significant variety of DQC hardware, ranging from those with a limited number of communication qubits to those having substantial capability to generate remote entanglements simultaneously. QuComm, on the other hand, lacks a global viewpoint of inter-correlation among blocks, showing less adaptation to the changes in EPR capacity, eventually leading to inefficient routing decisions and an increase in communication costs.

\begin{figure}[!t] 
    \centering
    \includegraphics[width=\columnwidth]{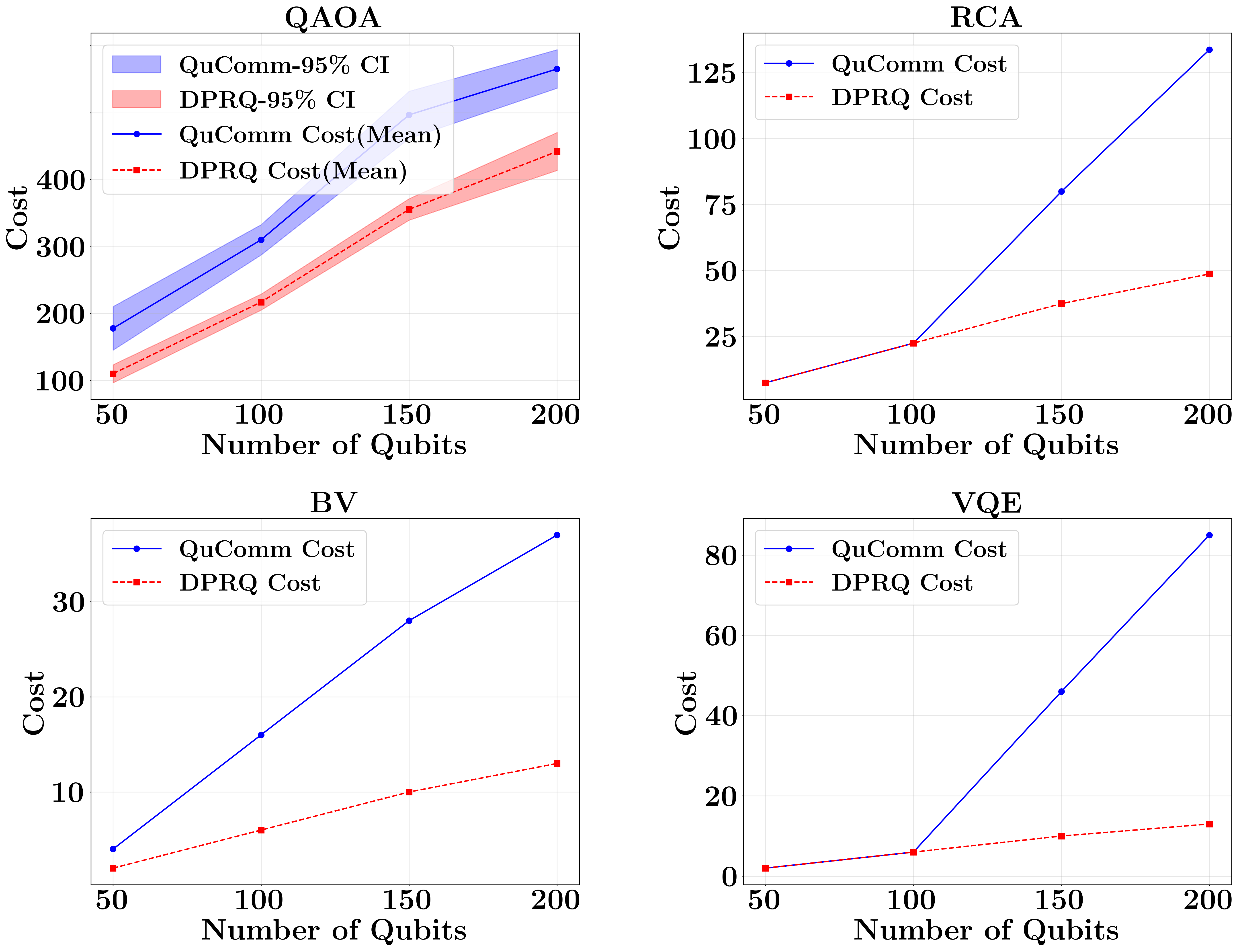}
    \caption{Cost comparison of DPRQ and QuComm on the 4 benchmark circuits as total number of qubits increases from 50 to 200. The EPR capacity is 3 and the number of qubits per node is 20.}
    \label{fig:results-qubit-count}
    \Description{This figure contains four graphs, each containing two line plots. The red line plot shows the EPR cost for DPRQ and blue line plot shows the EPR cost of routing using the baseline as the number of qubits increases from 50 to 200. The top-left graph displays data for QAOA, the top-right graph displays data for RCA, the bottom-left graph displays data for BV, and the bottom-right graph displays data for VQE. Additionally, in the QAOA graph, there is a 95\% confidence interval shown for both DPRQ and the baseline.}
\end{figure}

With varying circuit widths, DPRQ offers significantly higher reduction in inter-node communication compared to QuComm in all four benchmark circuits, as demonstrated in Figure~\ref{fig:results-qubit-count}. In RCA and VQE circuits, although QuComm performs equally well as DPRQ when the size of the DQC network is at most five nodes, its advantage disappears with the expansion of the network. 

\begin{figure}[!t]
    \centering
    \includegraphics[width=\columnwidth]{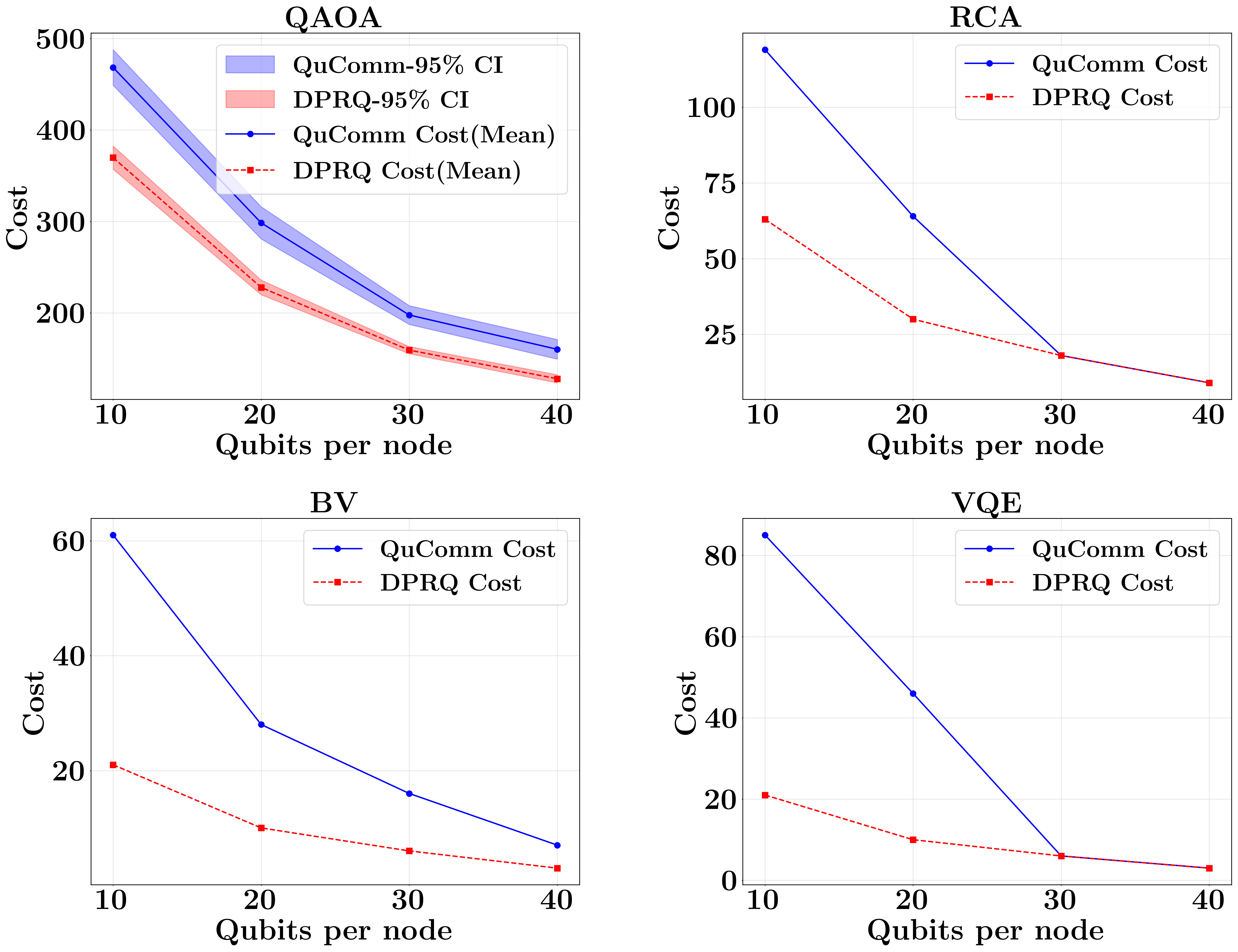}
    \caption{Cost comparison of DPRQ and QuComm on the four benchmark circuits as number of qubits per node increases from 10 to 40. The EPR capacity is 3 and the total number of qubits is 150.}
    \label{fig:qub-per-node}
    \Description{This figure contains four graphs, each containing two line plots. The red line plot shows the EPR cost for DPRQ and blue line plot shows the EPR cost of routing using the baseline as the number of qubits per node increases from 10 to 40. The top-left graph displays data for QAOA, the top-right graph displays data for RCA, the bottom-left graph displays data for BV, and the bottom-right graph displays data for VQE. Additionally, in the QAOA graph, there is a 95\% confidence interval shown for both DPRQ and the baseline.}
\end{figure}

With an increase in the number of data qubits per node, a huge number of collective communication blocks are no longer needed, diminishing the importance of the dynamic programming-based approach. This can be observed in Figure~\ref{fig:qub-per-node}. However, the advantage of DPRQ increases for a lesser number of data qubits per node and the results indicate a clear advantage of DPRQ over QuComm, which consistently increases with an increase in the number of nodes in the DQC network.

\subsection{Effect of Network Topologies}
\begin{figure}[!t]
    \centering
    \includegraphics[scale=0.25]{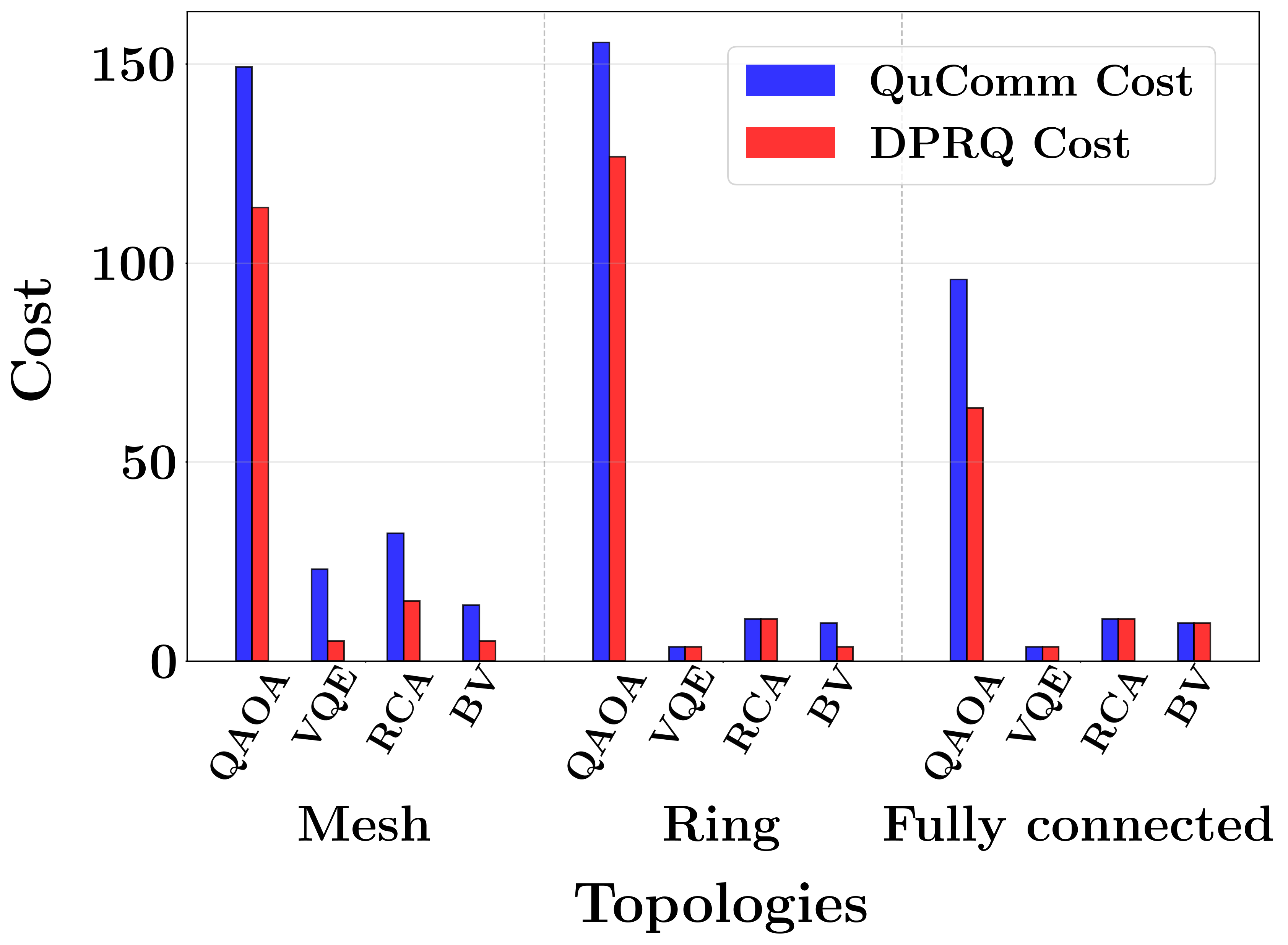}
    \caption{Cost comparison of DPRQ and QuComm for 3 different network topologies on the four benchmark circuits. The EPR capacity is 3, total number of qubits is 150 and the number of qubits per node is 20.}
    \label{fig:results-topology}
    \Description{This figure describes the EPR cost comparison of DPRQ and the baseline for three different network topologies and all the benchmark circuits. The first two bars describe the costs of the baseline and DPRQ respectively for QAOA circuit on mesh grid topology. The first bar is blue (for baseline) and the second bar is red (for DPRQ). The next two bars show similar information for VQE circuit on mesh grid topology. The next two bars are for RCA and the next two bars describe the costs of BV circuit on mesh grid topology. Then, the next two bars describe the costs of the baseline and DPRQ respectively for QAOA circuit on ring topology. The next two bars show similar information for VQE circuit on ring topology. The next two bars are for RCA and the next two bars describe the costs of BV circuit on ring topology. The next two bars describe the costs of the baseline and DPRQ respectively for QAOA circuit on fully-connected topology. The next two bars show similar information for VQE circuit on fully-connected topology. The next two bars are for RCA and the next two bars describe the costs of BV circuit on fully-connected topology.}
\end{figure}

To study the adaptability of DPRQ on a range of DQC network topologies, we used three different topologies to evaluate the performance of DPRQ, namely mesh-grid topology, ring topology, and a fully-connected topology. The results of the experiments performed using the default parameters are demonstrated in Figure~\ref{fig:results-topology}.
In all three topologies, DPRQ performs better than or equal to QuComm for all 4 benchmark circuits. Moreover, the benefit of DPRQ diminishes for a fully-connected DQC network topology. This is due to the fact that each node has an established communication link with every other node, which diminishes the importance of an improved aggregator node selection by DPRQ. However, it is clear that DPRQ is still dominant compared to QuComm due to the fact that in none of the cases, QuComm outperforms DPRQ. These results show that DPRQ is resilient to change in network topology, which makes it more scalable and significantly more adaptable to a large number of distributed systems using a variety of different network topologies.


%% file: conclusion.tex
\section{Conclusion}
\label{sec:conclusion}
This paper introduces DPRQ, a qubit routing algorithm that reduces the number of EPR pairs required for inter-node communication. DPRQ devises an intelligent strategy for intra-block as well as inter-block qubit routing once the circuit is divided into multiple collective communication blocks. Specifically, since inter-node communication within a block is highly dependent on the initial qubit layout, DPRQ employs a dynamic programming-based approach to explore multiple initial layouts, finally settling on the one that offers the most reduction in inter-node communication.

The results demonstrate that DPRQ brings a significant reduction in inter-node communication for a variety of DQC circuit configurations. Moreover, DPRQ's advantage grows as the number of nodes in a DQC network increases. This makes it much more adaptable for current as well as future DQC networks with a higher number of quantum processors. DPRQ has been shown to be resilient to changes in collective communication block patterns as well as network topologies.

With the enhanced EPR cost
calculation and dynamic program\-ming-based circuit optimization, DPRQ provides a strong qubit routing framework for inter-node cost reduction compared to the greedy routing approach employed by QuComm. We used 80 different configurations of the four benchmark circuits for evaluation and the results demonstrate that DPRQ offers an average reduction of 24.40\%  and a maximum reduction of 85.06\% in the EPR cost. Moreover, for a single circuit configuration, DPRQ offers a maximum average reduction of 48.54\% in four different benchmark circuits. 
The configuration in the discussion is as follows: Number of qubits = 200, number of qubits per node = 20, and EPR capacity = 2. This shows that for a vast DQC network having a limited number of communication qubits, DPRQ offers a significant reduction in inter-node communication regardless of the benchmark circuit used.


\begin{acks}
    Yu's research was supported in part by NSF grant 2350152. The information reported herein does not reflect the position or the policy of the funding agencies.
\end{acks}

%% file: reference.bib
@Article{Chatterjee2022,
author={Chatterjee, Turbasu
and Das, Arnav
and Mohtashim, Shah Ishmam
and Saha, Amit
and Chakrabarti, Amlan},
title={Qurzon: A Prototype for a Divide and Conquer-Based Quantum Compiler for Distributed Quantum Systems},
journal={SN Computer Science},
year={2022},
month={Jun},
day={10},
volume={3},
number={4},
pages={323},
issn={2661-8907},
doi={10.1007/s42979-022-01207-9},
url={https://doi.org/10.1007/s42979-022-01207-9}
}

@misc{javadiabhari2024quantumcomputingqiskit,
      title={Quantum computing with Qiskit}, 
      author={Ali Javadi-Abhari and Matthew Treinish and Kevin Krsulich and Christopher J. Wood and Jake Lishman and Julien Gacon and Simon Martiel and Paul D. Nation and Lev S. Bishop and Andrew W. Cross and Blake R. Johnson and Jay M. Gambetta},
      year={2024},
      eprint={2405.08810},
      archivePrefix={arXiv},
      primaryClass={quant-ph},
      url={https://arxiv.org/abs/2405.08810}, 
}

@misc{farhi2014quantumapproximateoptimizationalgorithm,
      title={A Quantum Approximate Optimization Algorithm}, 
      author={Edward Farhi and Jeffrey Goldstone and Sam Gutmann},
      year={2014},
      eprint={1411.4028},
      archivePrefix={arXiv},
      primaryClass={quant-ph},
      url={https://arxiv.org/abs/1411.4028}, 
}

@article{Vedral1996,
   author = {Vlatko Vedral and Adriano Barenco and Artur Ekert},
   doi = {10.1103/PhysRevA.54.147},
   issn = {1050-2947},
   issue = {1},
   journal = {Physical Review A},
   month = {7},
   pages = {147-153},
   title = {Quantum networks for elementary arithmetic operations},
   volume = {54},
   year = {1996}
}

@article{Peruzzo2014,
   author = {Alberto Peruzzo and Jarrod McClean and Peter Shadbolt and Man-Hong Yung and Xiao-Qi Zhou and Peter J. Love and Alán Aspuru-Guzik and Jeremy L. O’Brien},
   doi = {10.1038/ncomms5213},
   issn = {2041-1723},
   issue = {1},
   journal = {Nature Communications},
   month = {7},
   pages = {4213},
   title = {A variational eigenvalue solver on a photonic quantum processor},
   volume = {5},
   year = {2014}
}

@article{Bernstein1997,
   author = {Ethan Bernstein and Umesh Vazirani},
   doi = {10.1137/S0097539796300921},
   issn = {0097-5397},
   issue = {5},
   journal = {SIAM Journal on Computing},
   month = {10},
   pages = {1411-1473},
   title = {Quantum Complexity Theory},
   volume = {26},
   year = {1997}
}

@article{Ruf2021,
   author = {Maximilian Ruf and Noel H. Wan and Hyeongrak Choi and Dirk Englund and Ronald Hanson},
   doi = {10.1063/5.0056534},
   issn = {0021-8979},
   issue = {7},
   journal = {Journal of Applied Physics},
   month = {8},
   title = {Quantum networks based on color centers in diamond},
   volume = {130},
   year = {2021}
}

@article{Pompili2021,
   author = {M. Pompili and S. L. N. Hermans and S. Baier and H. K. C. Beukers and P. C. Humphreys and R. N. Schouten and R. F. L. Vermeulen and M. J. Tiggelman and L. dos Santos Martins and B. Dirkse and S. Wehner and R. Hanson},
   doi = {10.1126/science.abg1919},
   issn = {0036-8075},
   issue = {6539},
   journal = {Science},
   month = {4},
   pages = {259-264},
   title = {Realization of a multinode quantum network of remote solid-state qubits},
   volume = {372},
   year = {2021}
}

@article{Zomorodi-Moghadam2018,
   author = {Mariam Zomorodi-Moghadam and Mahboobeh Houshmand and Monireh Houshmand},
   doi = {10.1007/s10773-017-3618-x},
   issn = {0020-7748},
   issue = {3},
   journal = {International Journal of Theoretical Physics},
   month = {3},
   pages = {848-861},
   title = {Optimizing Teleportation Cost in Distributed Quantum Circuits},
   volume = {57},
   year = {2018}
}

@inproceedings{H_ner_2021, 
   series={SC ’21},
   title={Distributed quantum computing with QMPI},
   url={http://dx.doi.org/10.1145/3458817.3476172},
   DOI={10.1145/3458817.3476172},
   booktitle={Proceedings of the International Conference for High Performance Computing, Networking, Storage and Analysis},
   publisher={ACM},
   author={Häner, Thomas and Steiger, Damian S. and Hoefler, Torsten and Troyer, Matthias},
   year={2021},
   month=Nov, pages={1–13},
   collection={SC ’21} 
}

@article{Ferrari2021,
   author = {Davide Ferrari and Angela Sara Cacciapuoti and Michele Amoretti and Marcello Caleffi},
   doi = {10.1109/TQE.2021.3053921},
   issn = {2689-1808},
   journal = {IEEE Transactions on Quantum Engineering},
   pages = {1-20},
   title = {Compiler Design for Distributed Quantum Computing},
   volume = {2},
   year = {2021}
}

@article{DiAdamo2021,
   author = {Stephen DiAdamo and Marco Ghibaudi and James Cruise},
   doi = {10.1109/TQE.2021.3057908},
   issn = {2689-1808},
   journal = {IEEE Transactions on Quantum Engineering},
   pages = {1-21},
   title = {Distributed Quantum Computing and Network Control for Accelerated VQE},
   volume = {2},
   year = {2021}
}

@article{Davarzani2020,
   author = {Zohreh Davarzani and Mariam Zomorodi-Moghadam and Mahboobeh Houshmand and Mostafa Nouri-baygi},
   doi = {10.1007/s11128-020-02871-7},
   issn = {1570-0755},
   issue = {10},
   journal = {Quantum Information Processing},
   month = {10},
   pages = {360},
   title = {A dynamic programming approach for distributing quantum circuits by bipartite graphs},
   volume = {19},
   year = {2020}
}

@article{Daei2020,
   author = {Omid Daei and Keivan Navi and Mariam Zomorodi-Moghadam},
   doi = {10.1007/s10773-020-04633-8},
   issn = {0020-7748},
   issue = {12},
   journal = {International Journal of Theoretical Physics},
   month = {12},
   pages = {3804-3820},
   title = {Optimized Quantum Circuit Partitioning},
   volume = {59},
   year = {2020}
}

@article{Dadkhah2022,
   author = {Davood Dadkhah and Mariam Zomorodi and Seyed Ebrahim Hosseini and Pawel Plawiak and Xujuan Zhou},
   doi = {10.1109/ACCESS.2022.3186485},
   issn = {2169-3536},
   journal = {IEEE Access},
   pages = {70329-70341},
   title = {Reordering and Partitioning of Distributed Quantum Circuits},
   volume = {10},
   year = {2022}
}

@article{10.1098/rspa.2012.0686,
    author = {Beals, Robert and Brierley, Stephen and Gray, Oliver and Harrow, Aram W. and Kutin, Samuel and Linden, Noah and Shepherd, Dan and Stather, Mark},
    title = {Efficient distributed quantum computing},
    journal = {Proceedings of the Royal Society A: Mathematical, Physical and Engineering Sciences},
    volume = {469},
    number = {2153},
    pages = {20120686},
    year = {2013},
    month = {05},
    issn = {1364-5021},
    doi = {10.1098/rspa.2012.0686},
    url = {https://doi.org/10.1098/rspa.2012.0686},
    eprint = {https://royalsocietypublishing.org/rspa/article-pdf/doi/10.1098/rspa.2012.0686/834214/rspa.2012.0686.pdf},
}

@article{Andrs-Martnez2019,
   author = {Pablo Andrés-Martínez and Chris Heunen},
   doi = {10.1103/PhysRevA.100.032308},
   issn = {2469-9926},
   issue = {3},
   journal = {Physical Review A},
   month = {9},
   pages = {032308},
   title = {Automated distribution of quantum circuits via hypergraph partitioning},
   volume = {100},
   year = {2019}
}

@article{Hermans2022,
   author = {S. L. N. Hermans and M. Pompili and H. K. C. Beukers and S. Baier and J. Borregaard and R. Hanson},
   doi = {10.1038/s41586-022-04697-y},
   issn = {0028-0836},
   issue = {7911},
   journal = {Nature},
   month = {5},
   pages = {663-668},
   title = {Qubit teleportation between non-neighbouring nodes in a quantum network},
   volume = {605},
   year = {2022}
}

@article{Daiss2021,
   author = {Severin Daiss and Stefan Langenfeld and Stephan Welte and Emanuele Distante and Philip Thomas and Lukas Hartung and Olivier Morin and Gerhard Rempe},
   doi = {10.1126/science.abe3150},
   issn = {0036-8075},
   issue = {6529},
   journal = {Science},
   month = {2},
   pages = {614-617},
   title = {A quantum-logic gate between distant quantum-network modules},
   volume = {371},
   year = {2021}
}

@inproceedings{10.1145/3613424.3614253,
author = {Wu, Anbang and Ding, Yufei and Li, Ang},
title = {QuComm: Optimizing Collective Communication for Distributed Quantum Computing},
year = {2023},
isbn = {9798400703294},
publisher = {Association for Computing Machinery},
address = {New York, NY, USA},
url = {https://doi.org/10.1145/3613424.3614253},
doi = {10.1145/3613424.3614253},
booktitle = {Proceedings of the 56th Annual IEEE/ACM International Symposium on Microarchitecture},
pages = {479–493},
numpages = {15},
location = {Toronto, ON, Canada},
series = {MICRO '23}
}

@misc{wu2022autocommframeworkenablingefficient,
      title={AutoComm: A Framework for Enabling Efficient Communication in Distributed Quantum Programs}, 
      author={Anbang Wu and Hezi Zhang and Gushu Li and Alireza Shabani and Yuan Xie and Yufei Ding},
      year={2022},
      eprint={2207.11674},
      archivePrefix={arXiv},
      primaryClass={quant-ph},
      url={https://arxiv.org/abs/2207.11674}, 
}

@inproceedings{10.1109/INFOCOM.2018.8486419,
author = {Leconte, Mathieu and Destounis, Apostolos and Paschos, Georgios},
title = {Traffic Engineering with Precomputed Pathbooks},
year = {2018},
publisher = {IEEE Press},
url = {https://doi.org/10.1109/INFOCOM.2018.8486419},
doi = {10.1109/INFOCOM.2018.8486419},
booktitle = {IEEE INFOCOM 2018 - IEEE Conference on Computer Communications},
pages = {234–242},
numpages = {9},
location = {Honolulu, HI, USA}
}
